\documentclass[prb,reprint,amssymb]{revtex4-2}

\usepackage{amsmath}  
\usepackage{amsfonts}
\usepackage{graphicx}
\usepackage{hyperref}
\hypersetup{
    colorlinks=true,
	citecolor=blue,
    linkcolor=blue,
	urlcolor = blue
}

\begin{document}

\title{Separation of quadrupole, spin, and charge across the magnetic phases of a one-dimensional interacting spin-1 gas}

\author{Felipe Reyes-Osorio}
\author{Karen Rodr\'iguez-Ram\'irez}
\email{karem.c.rodriguez@correounivalle.edu.co}
\affiliation{Departamento de f\'isica, Universidad del Valle, Cali, Colombia, 760032}

\date{\today}

\begin{abstract}
We study the low-energy collective properties of a 1D spin-1 Bose gas using bosonization. After giving an overview of the technique, emphasizing the physical aspects, we apply it to the $S=1$ Bose-Hubbard Hamiltonian and find a novel separation of the quadrupole-spin-charge sectors, confirmed by time-MPS numerical simulations. Additionally, through the single particle spectrum, we show the existence of the superfluid-Mott insulator transition and the point at which the physics are described by a Heisenberg-like Hamiltonian. The magnetic phase diagrams are found for both the superfluid and insulating regimes; the latter is determined by decomposing the complete Heisenberg bilinear-biquadratic Hamiltonian, which describes the Mott insulator, into simpler, effective Hamiltonians. This allows us to keep our methods flexible and  transferable to other interesting interacting condensed matter systems.
\end{abstract}

\maketitle

\section{Introduction.}\label{secIntro}

Quantum simulation has developed rapidly in the past twenty years due to the emergence of practical experimental cooling and trapping techniques \cite{chu, cohen, phillips, bouyer2000}. Many milestones have been achieved in ultracold atoms and quantum simulators, such as the realization of a Bose-Einstein condensate in the lab, and the direct observation of the superfluid-Mott insulator (SF-MI) transition \cite{Greiner2008, Bloch2005, Grimm2000, Gross995, Greiner2002}. These exemplify the ability to simulate a variety of condensed matter models in impurity-free optical lattices orders of magnitude larger than natural crystalline structures. However, while interaction parameters can be finely tuned, \textit{e.g.} through Feshbach resonances \cite{nagerl2008, bloch2009}, simulating magnetic properties using neutral atoms has been a challenge met with innovative methods for creating artificial gauge fields \cite{aidelsburger2018, galitski2019}. These range from physically shaking or rotating the lattice \cite{Zwierlein2005, struck2012}, to laser induced tunneling \cite{Lin2009}, to the microwave ac Zeeman effect \cite{gerbier2006}. More exotic setups couple to the internal degrees of freedom of the atoms and simulate a wide array of phenomena, from additional synthetic dimensions \cite{stuhl2015} to topological many-body states \cite{cooper2013}. Since quantum simulators aimed at these exciting prospects heavily depend on coupling to the atomic spin it becomes necessary to understand the behavior of spinful systems and their existing magnetic properties.

The study of higher-spin systems and $SU(N)$ \cite{manmana2011} magnetism has developed synergistically alongside ultracold atoms and quantum simulators, expanding upon the rich knowledge of spin-$\frac{1}{2}$ magnetism, and yielding ever-more sophisticated experiments \cite{Gorshkov2010}. In particular, the study of spin-1 systems is of great relevance because most common alkali isotopes used to populate optical lattices have a spin-1 hyperfine ground state \cite{deuretzbacher2008}. Additionally, these systems exhibit novel magnetic behavior due to the importance of the quadrupole tensor and the new interactions that this quantity allows for \cite{bulhakov, peletminski, toth2012, moura2011, papanicolau1988, chubukov1991, Podolsky2005, kolezhuk2008}. Ordering of these new degrees of freedom leads to nematic and topological behavior absent from conventional spin-$\frac{1}{2}$ 1D systems. Although much research has been devoted to systems with explicit quadrupole-quadrupole interactions, quadrupole degrees of freedom are an intrinsic feature of spin-1 particles and emerge even when considering standard spinful contact pseudopotentials, as we will later show. 

\begin{figure}
\centering
\includegraphics[width=\linewidth]{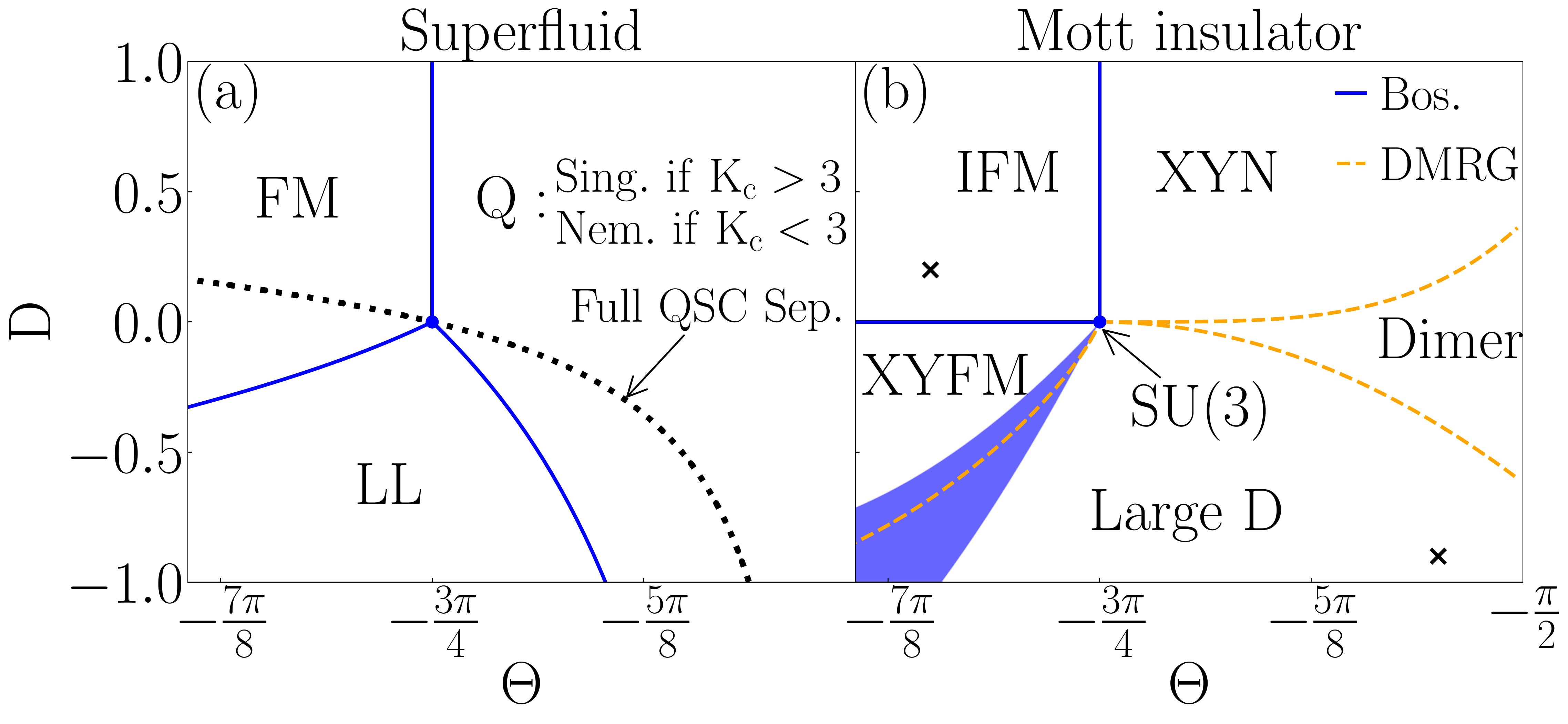}
\caption{\label{figPhaseDiag} Magnetic phases of the 1D spin-1 Bose gas. The parameters $D$ and $\Theta$ control the quadratic Zeeman field and the interaction strengths respectively, and are explained with more detail in later sections. (a) are the phases within the SF regime, while (b) the ones within the MI. In blue, the transitions determined analytically with bosonization; in orange dashed line, the remaining phase transitions known from previous DMRG study (colors online) \cite{Rodriguez_2011}. Black crosses in (b) are the simulated parameters in Sec. \ref{secHBB}.}
\end{figure}

\begin{figure}
\centering
\includegraphics[width=\linewidth]{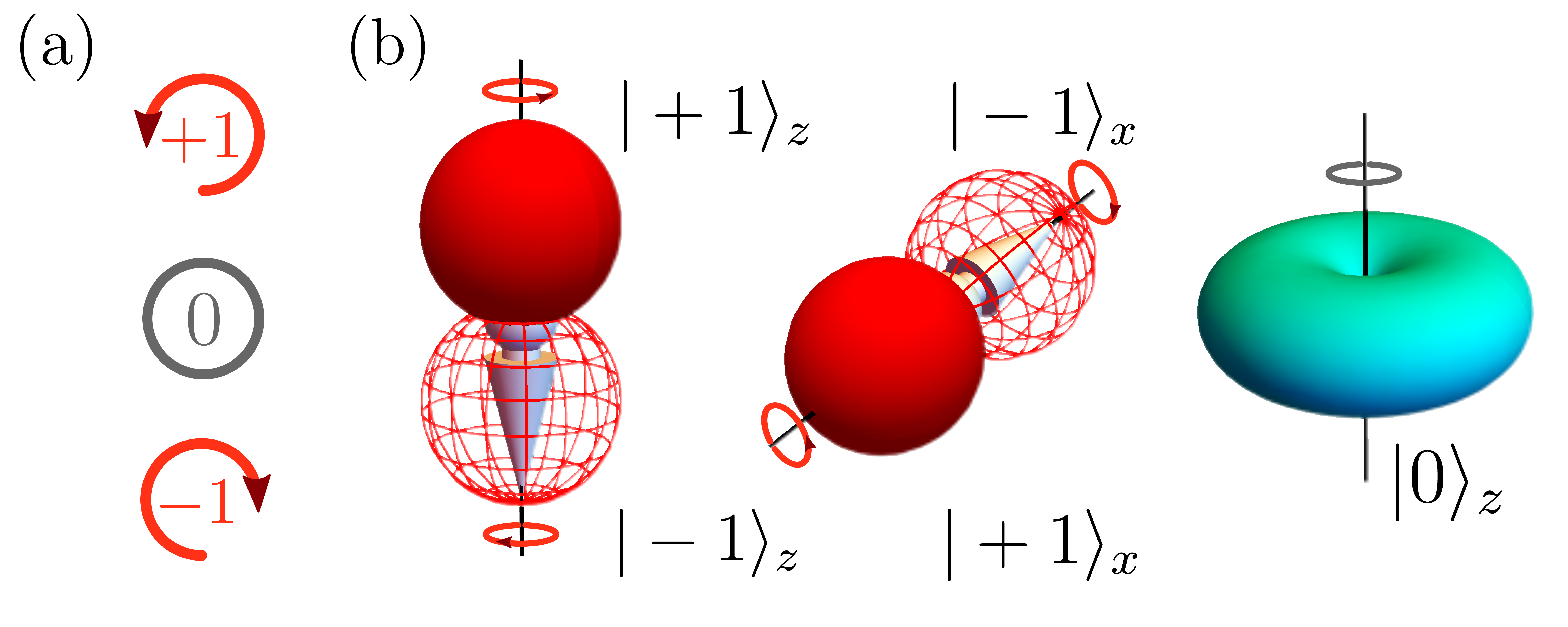}
\caption{\label{figStates} Graphical representation of spin-1 states. (a) shows our rotation convention to denote the three projections. (b) shows specific states projected onto the coherent spin basis; magnetizable spinor states can have an assigned direction, whereas non-magnetizable quadrupolar states only have a preferred nematic axis.}
\end{figure} 

The new quadrupolar effects found in spin-1 systems are significantly enhaced when considering 1D systems, another topic with decades worth of theory brought back to the lab by optical lattices \cite{Franchini2017, fallani2004, Schmid2006}. One of the defining properties of 1D systems is their tendency towards collective behavior due to the inevitability of iteractions between neighboring particles, leading to global excitations \cite{Giamarchi2016, giamarchi}. Consequently, at low energies, descriptions are often in terms of emergent quasiparticles of an effective field theory, leading to analytical approaches such as bosonization.

In this paper, we study a gas of 1D spin-1 bosons subject to an optical lattice, and apply bosonization in order to extract physical information from the low-energy regime without resorting to mean-field approximations. Therefore, after giving an overview to bosonization in Sec. \ref{secOverBos}, we establish the important physical properties of spin-1 systems, and formulate a many-body Hamiltonian in Sec. \ref{secBH}. After applying bosonization, we identify the relevant degrees of freedom of the system, and their novel separation. Then, in Sec. \ref{secSF}, we obtain the magnetic phase diagram for the superfluid (SF) regime, shown in Fig. \ref{figPhaseDiag}(a), from our bosonized Hamiltonian. Then, we do the same for the Mott insulator (MI) regime in Sec. \ref{secHBB}, by first decomposing the system into simpler, effective Hamiltonians and bosonizing; the phase diagram is shown in Fig. \ref{figPhaseDiag}(b). Finally, Sec. \ref{secConclusion} contains conclusions to our work.

\section{Overview of bosonization} \label{secOverBos}

There are plenty of reviews of bosonization \cite{giamarchi, von_Delft_1998,snchal, gogolin, giuliani2005, miranda, affleckFields, cazalilla,solyom1979,Haldane1981}; here we wish to briefly go over the technique, highlighting important physical connections and establish the main equations that will be used in the rest of the text.

Bosonization exploits the strong tendency towards collective behavior that 1D systems exhibit due to their dimensionality; at low energies, this is manifested in the emergent quasiparticles of the system. In fermionic systems, the low-energy behavior is that of a Luttinger liquid (LL) \cite{tomonaga, luttinger}, whose linear dispersion matches the dispersion of massive fermions around the Fermi momentum $k_F$. As such, real fermions $\hat c$ can be mapped into the chiral fermions of the LL according to 
\begin{equation}
\hat c(x) = e^{ik_Fx}\hat\psi_R(x) + e^{-ik_Fx}\hat\psi_L(x), \label{eqPosMap}
\end{equation}
where $\hat \psi_\alpha(k)$ annihilates a chiral fermion with momentum $k$ in the $\alpha=R,L$ branch, and $\hat \psi_\alpha(x)$ does the same in position $x$. The map shows that a localized fermion is a superposition of long-wavelength chiral oscillations on top of the Fermi sea \cite{cazalilla}. The map is made rigorous by defining a normal-ordering of operators with respect to the Fermi surface, that matches the infinitely-occupied ground state of the LL. 

The quasiparticles of the LL, and consequently of low-energy fermions, are bosonic superpositions of particle-hole excitations, $\hat b_p = -i\sqrt{2\pi/|p|}(\Theta(p)\hat J_R(p) - \Theta(-p)\hat J_L(-p))$, where $\hat J_\alpha(p) = \frac{1}{\sqrt L} \sum_q {\hat \psi_\alpha^\dagger(q-\alpha p)\hat \psi_\alpha(q)}$ and the boson commutation relations are due to the infinitely-occupied vacuum \cite{mattis}. $\Theta(p)$ is the unit step function. These $\hat b^\dagger_p/\hat b_p$ create/annihilate collective superpositions of particle-hole excitations with well-defined momentum $p$ in the $R(L)$ branch for $p>0 \ (p<0)$. The bosonic modes lead to the formulation of a scalar field $\hat \Phi$ and its canonical momentum $\hat \Pi$ \cite{schwartzQFT}
\begin{eqnarray}
\hat \Phi (x) &=& \sum_{p\neq 0} \frac{1}{\sqrt{2\pi |p|}}e^{ipx}(\hat b_p + \hat b_{-p}^\dagger) \\
\hat \Pi (x) &=& -i\sum_{p\neq 0} \sqrt{\frac{|p|}{2\pi}}e^{ipx}(\hat b_p - \hat b_{-p}^\dagger), \nonumber
\end{eqnarray}
which are connected to the chiral fermions through the fundamental bosonization identity \cite{von_Delft_1998}
\begin{equation}
\hat \psi_\alpha(x) = \frac{1}{\sqrt{2\pi\delta}}e^{\alpha i 2\sqrt{\pi} \hat{\phi}_\alpha(x)}, \label{eqBosIdF}
\end{equation}
where $\delta$ is the cutoff length, and the chiral fields $\hat \phi_\alpha$ are the contributions of the $\alpha$ branch to the scalar field $\hat \Phi$. 

The bosonic formulation of the LL, and thus of real interacting fermions, is summarized in the Hamiltonians
\begin{eqnarray}
\hat H &=& u\sum_{p\neq 0} |p|\hat{\tilde b}^\dagger_p \hat{\tilde b}_p, \label{eqLuttBos} \\
\hat H &=& \frac{u}{2}\int dx \left(\frac{1}{K}(\partial_x \hat \Phi)^2 +K \hat \Pi^2\right), \label{eqLuttFields}
\end{eqnarray}
where $\hat{\tilde b}^\dagger_p / \hat{\tilde b}_p$ are related to the original bosons through a unitary transformation, $u$ is the propagation velocity renormalized by forward-scattering interactions \cite{solyom1979}. The Luttinger parameter $K$ can be thought to hold all the information about the interaction, and is related to physical quantities such as the compressibility $\kappa \propto u/K$ and the heat capacity $c\propto uK$. For free fermions, $K = 1$ and $u = v_F$; repulsive fermions are characterized by $K<1$ and attractive fermions by $K>1$. It is important to note that only long-wavelength interactions, of the type $g_2$ and $g_4$ in the standard g-ology~\cite{solyom1979}, can renormalize $K$. However, more general interactions, like backscatering, take the form of more complex functions of exponentials due to the form of the fundamental bosonization identitity [Eq.~\ref{eqBosIdF}]. Often, these show up as sine-Gordon terms like $\hat H_1 = g_1\int dx\ \cos(a\sqrt{\pi} \hat\Phi)$, where the real coefficient $a$ determines the scaling dimension \cite{giamarchi}. Finally, an imaginary $K$ indicates a break in the Luttinger liquid approximation, indicating a vanishing compressibility and we can again expect the system to order~\cite{schulz}.

Within bosonization, sine-Gordon terms are typically analyzed with standard renormalization group techniques~\cite{giamarchi}, and indicate an ordered phase when the corresponding field gets trapped in one of the minima. On the other hand, the main method to determine the phases of the system is to study the decay of the correlation functions. In 1D, this decay is algebraic, characterized by a scaling dimension $\gamma$ related to the Luttinger parameter $K$. Although a continuous symmetry cannot be spontaneously broken in 1D, through the correlation function we can find divergences in the susceptibilities, indicating a quasiordering in the system~\cite{giamarchi}. When multiple factors diverge, the dominant order as determined from $\gamma$ prevails.

In bosonic systems, bosonization is based on the low-energy hydrodynamic description, in terms of the density and phase. From the LL, we know that these (normal-ordered) variables are ${\hat \rho(x)} = \partial_x\hat \Phi/\sqrt\pi$, and $\partial_x \hat \theta = \hat \Pi$. Thus, the fundamental bosonization identity for bosons \cite{cazalilla, giamarchi} is given by
\begin{equation}
\hat a(x) = \sqrt{\bar \rho + \frac{1}{\pi}\partial_x\hat \Phi}\sum_{n=0,\pm 1} e^{i2n(\pi\bar\rho x + \sqrt\pi\hat\Phi)} e^{i\sqrt{\pi}\hat \theta}, \label{eqBosIdB}
\end{equation}
where $\bar \rho$ is the average ground state density, and the exclusion of odd harmonics determines the bosonic relations. With Eq. \ref{eqBosIdB}, it is possible to rewrite repulsive bosons as a LL, {\it i.e,} as the Hamiltonian of Eq. \ref{eqLuttFields}. In this case, $K = 1$ corresponds to hard-core bosons, retrieving the well-known relation between a Tonks-Girardeau gas and free fermions \cite{girardeau1960}. For free bosons, $K = \infty$, the compressibility $\kappa$ vanishes, bosons collapse or condensate, and the system is a true superfluid. Attractive bosons yield $K\in \mathbb{C}$, and are not a LL; this regime can be interpreted as having a completely saturated density \cite{schulz}. Because free bosons have $K = \infty$, we cannot perturb around them, as we can for free fermions, so additional techniques are needed for quantitative descriptions \cite{giamarchi, cazalilla}. However, on its own, bosonization can extract meaningful physical information that qualitatively describes the system well.

\section{Spin-1 bosons}\label{secBH}

Before defining a Hamiltonian for spin-1 particles, let us determine the relevant operators for such a system. Since a spin-1 system can be at most symmetric under $SU(3)$, its generators $\hat \lambda^\alpha$ form a maximal set of possibly relevant operators, where $\alpha=1...8$. These generators can be divided into the spin vector $\hat{\mathbf{S}}_i = (\hat \lambda_i^7, -\hat \lambda_i^5, \hat \lambda_i^2)$, and the five independent components of the quadrupolar tensor $\hat{\mathbf{Q}}_i = -(\hat \lambda_i^1,\hat \lambda_i^3,\hat \lambda_i^4,\hat \lambda_i^6,-\hat \lambda_i^8)$ \cite{bulhakov, peletminski, toth2012}; the two commuting generators,  
\begin{equation}
\lambda^2 = \begin{pmatrix}
1 & 0 & 0 \\
0 & 0 & 0 \\
0 & 0 & -1 \\
\end{pmatrix} \qquad
\lambda^8 = \frac{1}{\sqrt 3}\begin{pmatrix}
1 & 0 & 0 \\
0 & -2 & 0 \\
0 & 0 & 1 \\
\end{pmatrix},
\end{equation} are the spin and quadrupole densities and thus, we expect the magnetic sector of any spin-1 system to be describable as a combination of these degrees of freedom.

\begin{figure}
\centering
\includegraphics[width=\linewidth]{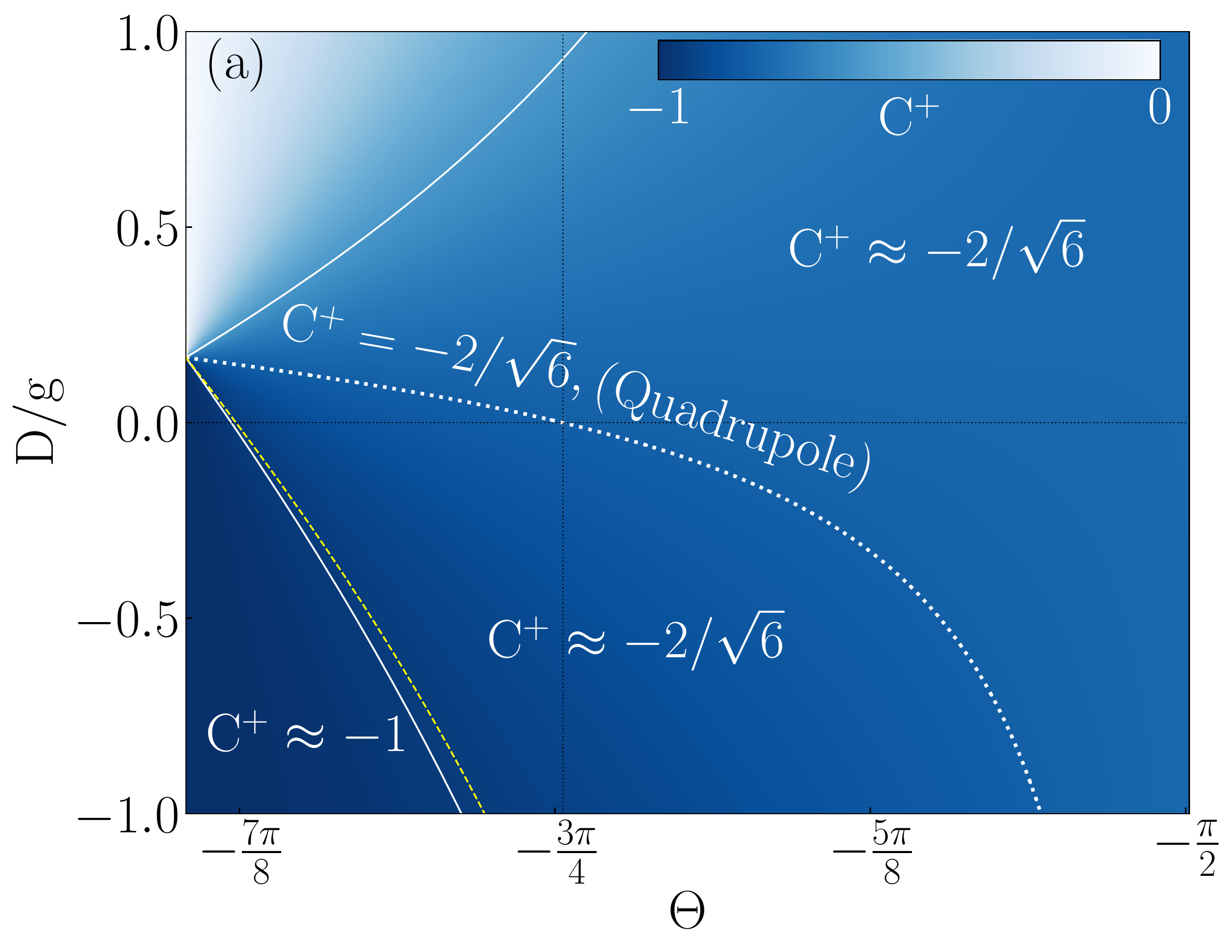}
\caption{\label{figSeparation} Coefficient of the $m=0$ component in the field $\hat \Phi^+$. Along the dotted line, $\hat \Phi^+ = \hat \Phi_q$, and complete QSC separation is achieved. Inside the white lines, the quadrupole and charge sectors have a 95 \% separation. Within the yellow dashed line, the $m=0$ projection separates from an effective spin-$\frac{1}{2}$ particle.}
\end{figure}

Having found the relevant degrees of freedom, we classify the possible states using by projecting onto the spin coherent states \cite{lee2015}, defined by a polar and azimuthal angle. In this work, we use the spin coherent states only as a visualization tool, and refer to Ref.~\cite{lee2015} for more details. The states are plotted in Fig. \ref{figStates}: the magnetic projections $|\pm 1\rangle$ are seen to have a preferred direction and so can also be represented as arrows with clockwise or counterclokwise rotation; on the other hand, $|0\rangle$ has a preferred axis with no single direction, denominated director or nematic vector, and no rotation. The formers are magnetized and belong to the spin states, whereas the latter is a non-magnetized quadrupolar state. The magnetization or lack thereof is indicated by red and blue, respectively (colors online).

For two spin-1 particles, the general Hamiltonian in first quantization is $\hat H^{(2)} = \sum_{i=1}^2 \hat p^2_i/2m - D\sum_{i=1}^2 (\hat S_i^z )^2 + \hat V$, where $\hat p_i$ is the momentum operator of the $i$-th particle with mass $m$, $D$ is the coupling to a magnetic field through the quadratic Zeeman effect , and $\hat S^z_i$ is the $z$ spin-1 operator \cite{kawaguchi2012, ueda2013, ho1998, Ohmi-Machida1998}. The low-energy interaction between the particles, $\hat V$, can be modeled as a pseudopotential \cite{huang1957} and computed by adding their angular momentum and studying the structure of the coupled Hilbert space \cite{lewensteinBook}. This yields two allowed orthogonal interaction channels, with $S=0$ and $S=2$, so that $\hat V = \sum_S g_S \hat P_S \delta(x_1-x_2)$, where $\hat P_s$ is the projector to the interaction channel $S$ of strength $g_S$. Generalizing to a many-body second-quantized system, we obtain the so-called $S=1$ Bose-Hubbard Hamiltonian (BHH) \cite{imambekov2003}
\begin{eqnarray}
\hat{H}_{BH} &=&-t \sum_{i, \sigma}\left(\hat{a}_{i+1, \sigma}^{\dagger} \hat{a}_{i, \sigma}+\hat{a}_{i, \sigma}^{\dagger} \hat{a}_{i+1, \sigma}\right)-D \sum_{i, \sigma} \sigma^{2} \hat{n}_{i, \sigma} \nonumber \\
&+&\sum_{i}\bigg[\frac{g_{2}}{2}\Big(\hat{a}_{1}^{\dagger} \hat{a}_{1}^{\dagger} \hat{a}_{1} \hat{a}_{1}+\hat{a}_{-1}^{\dagger} \hat{a}_{-1}^{\dagger} \hat{a}_{-1} \hat{a}_{-1} \nonumber \\
&+&2 \hat{a}_{1}^{\dagger} \hat{a}_{0}^{\dagger} \hat{a}_{1} \hat{a}_{0} +2 \hat{a}_{-1}^{\dagger} \hat{a}_{0}^{\dagger} \hat{a}_{-1} \hat{a}_{0}\Big) \label{eqBHH}\\
&+&\frac{2 g_{2}+g_{0}}{6} \hat{a}_{0}^{\dagger} \hat{a}_{0}^{\dagger} \hat{a}_{0} \hat{a}_{0} +\frac{g_{2}+2 g_{0}}{3} \hat{a}_{1}^{\dagger} \hat{a}_{-1}^{\dagger} \hat{a}_{-1} \hat{a}_{1} \nonumber \\
&+&\frac{g_{2}-g_{0}}{3}\left(\hat{a}_{0}^{\dagger} \hat{a}_{0}^{\dagger} \hat{a}_{1} \hat{a}_{-1}+\hat{a}_{1}^{\dagger} \hat{a}_{-1}^{\dagger} \hat{a}_{0} \hat{a}_{0}\right)\bigg]_{i}, \nonumber 
\end{eqnarray}
where $\sigma=\pm 1,0$ for each magnetic projection, and we consider the relevant case of a balanced mixture.  Each term in the interaction denotes a possible process, of which there are two kinds: spin-preserving, in which the outgoing magnetic projections are the same as the incoming; and spin-changing, where the opposite is true, with strength proportional to $g_2-g_0$. This difference is naturally small in commonly used atomic vapors \cite{kempen2002,Widera2005,Widera2006}, and spin-changing interactions are further suppressed by the quadratic Zeeman field \cite{Klempt2009}. Additionally, the point $g_2= g_0$ exhibits full $SU(3)$ symmetry, and is further discussed in Sec. \ref{secHBB}; it has been shown that similar regimes of unbroken enhaced symmetry are robust against decoherence and dissipation \cite{Tomka2015}. Therefore, in what follows, we can treat the spin-changing contributions as small. 

Using Eq. \ref{eqBosIdB} and Eq. \ref{eqBHH}, we obtain the bosonized BHH
\begin{eqnarray}
\hat{H}&=&\hat{H}_{0}^{(1)}+\hat{H}_{0}^{(0)}+\hat{H}_{0}^{(-1)}\\
&+&\frac{1}{\pi} \int dx \ \partial_x\hat{\boldsymbol\Phi}^\dagger\left(\begin{array}{ccc}
\frac{g_{2}}{2} - D  & \frac{g_{2}}{2} & \frac{g_{2}+2 g_{0}}{6} \\
\frac{g_{2}}{2} & \frac{2 g_{2}+g_{0}}{6} & \frac{g_{2}}{2} \\
\frac{g_{2}+2 g_{0}}{6} & \frac{g_{2}}{2} & \frac{g_{2}}{2} - D
\end{array}\right)
\partial_x\hat{\boldsymbol\Phi} \nonumber \\
&+&\frac{2\Delta g}{3} \int dx \ \cos(\sqrt{6\pi}\hat\theta_q) \nonumber 
\end{eqnarray}
where the interaction matrix couples three scalar fields corresponding to magnetic projections: $\hat{\boldsymbol{\Phi}}^\dagger = (\hat \Phi_1, \hat \Phi_0, \hat \Phi_{-1})$. The quadratic Zeeman field is included in the interaction matrix, and the spin-changing term becomes the cosine of the quadrupole phase, $\hat \theta_q = \frac{1}{\sqrt 6 } (\hat \theta_1 -2\hat\theta_0 + \hat \theta_{-1})$, and is coupled through $\Delta g = g_2-g_0$. The free terms $\hat H^{(m)}_0$ are in the form of the Hamiltonian of Eq.~\ref{eqLuttFields} with $K=1$ and can be traced back to the hopping term, and the phase fields are defined through their relation with the momentum field $\partial_x\hat\theta_m = \hat\Pi_m$.

\begin{figure}
\centering
\includegraphics[width=\linewidth]{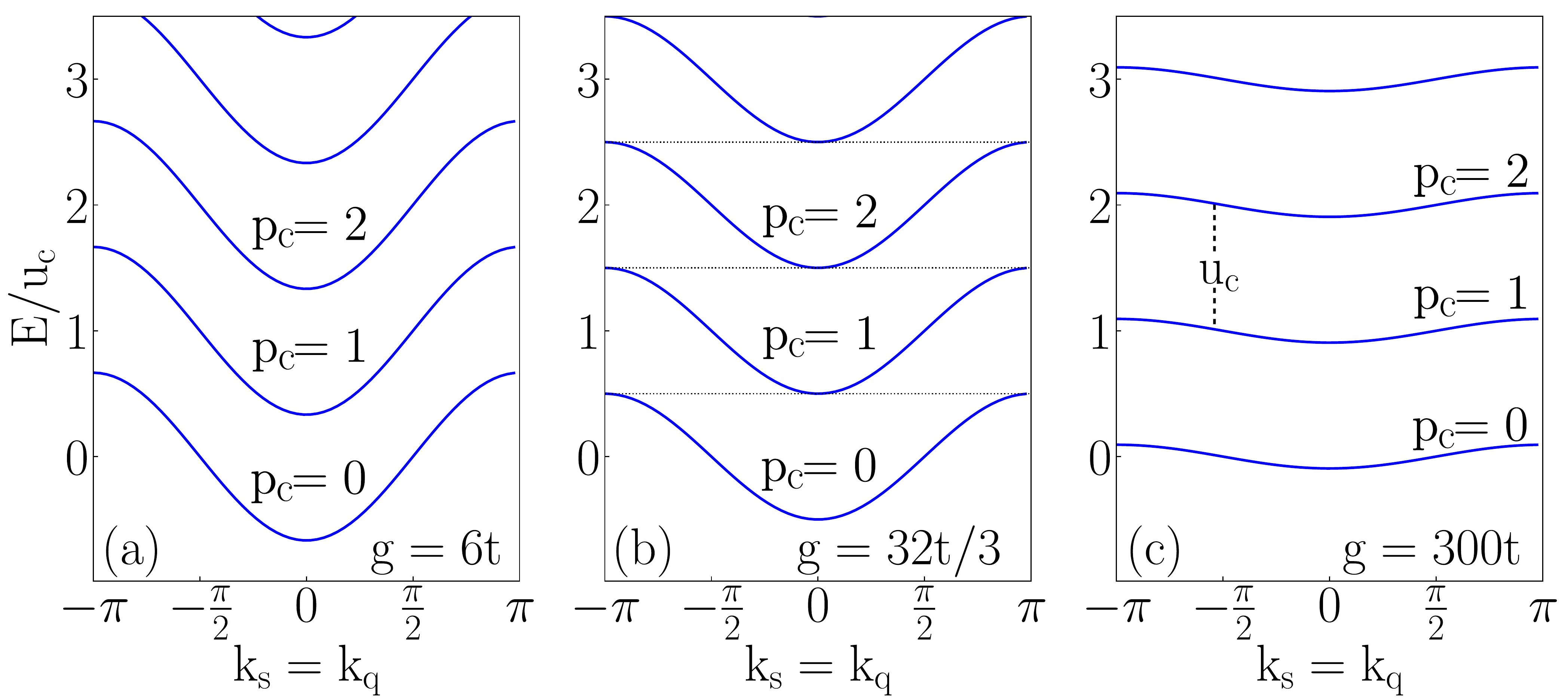}
\caption{\label{figSpectrum}Cross-section $k_s = k_q$ of the single-particle spectrum of the spin-1 gas at the symmetry point. As the interaction strength is increased from (a)-(c), the system becomes gapped in the charge sector, indicating the SF-MI phase transition. In this regime, the magnetic sectors become highly degenerate due to the absence of the quadratic Zeeman field.}
\end{figure}

The Hamiltonian is decoupled by transforming into the fields
\begin{eqnarray}
\hat \Phi_s &=& \frac{1}{\sqrt 2}(\hat \Phi_1 - \hat \Phi_{-1}) , \label{eqSectors} \\
\hat \Phi^\pm &=&  A^\pm\hat\Phi_1 + C^\pm\hat \Phi_0 + A^\pm\hat \Phi_{-1}, \nonumber
\end{eqnarray}
each corresponding to an independent sector of the system. Thus, the system separates into an independent magnetizable spin sector $\hat \Phi_s$, and two additional $\hat \Phi^\pm$. Since the sectors are orthonormal, and have only two independent coefficients, it suffices to study the behavior of $C^+$ to determine the nature of both contributions. At the symmetry point, $\Delta g = 0$, $D = 0$, and $C^+ = \frac{-2}{\sqrt 6}$, meaning
\begin{eqnarray}
\hat \Phi^+_{\rm sym}& =&  \hat \Phi_q = \frac{1}{\sqrt 6}(\hat \Phi_1 - 2 \hat\Phi_0 + \hat \Phi_{-1}), \\
\hat \Phi^-_{\rm sym} &= & \hat \Phi_c = \frac{1}{\sqrt 3}(\hat \Phi_1 + \hat\Phi_0 + \hat \Phi_{-1}). \nonumber
\end{eqnarray}
Thus, we find a novel separation of the quadrupole, spin, and charge (QSC) sectors. This underlines the importance of the quadrupole in spin-1 systems, since it emerges as a relevant sector even in the absence of explicit quadrupole-quadrupole interaction.

While spin-charge separation is a well known phenomenon in 1D spin-$\frac{1}{2}$ physics \cite{fuchs2005}, little is discussed about full QSC separation for spin-1 particles. Moreover, while a similar separation has been reported in the context of multispecies fermions \cite{ulbricht2010, azaria2009, Kundu2021} we show QSC separation in a bosonic system easily realized as well as appropriately characterize it in terms of the quadrupolar density, a physical quantity. Furthermore, to the best of our knowledge, QSC separation has not been examined away from the symmetry point, which we do in the following. Thus, let us reparametrize the interaction in terms of the angle $\Theta = \tan^{-1}((g_0+2g_2)/3g_0) \in [\tan^{-1}(1/3), -\pi/2)$, which takes values in the third quadrant in order to stay within the phase space of the original BHH. The value of $C^+$ is plotted in Fig. \ref{figSeparation}. The symmetry point is located at $\Theta = -3\pi/4$ and $D=0$, and full QSC separation is achieved along the dotted white line. However, to quantify the deviation away from full separation, we compute the projection of $\hat \Phi^+$ onto $\hat \Phi_q$, which shows that the decoupled fields vary very slowly around the symmetry point; in fact, for most of phase space (area bounded by the white lines of Fig. \ref{figSeparation}), $\hat \Phi^+ \cdot \hat \Phi_q > 0.95$ , meaning that the quadrupole and charge sectors are 95\% separated. In this sense, QSC separation is a robust property of the system, and approximating $\hat\Phi^+ \approx \hat\Phi_q$, and $\hat\Phi^- \approx \hat\Phi_c$ is equivalent to neglecting second order corrections to the interactions within sectors.

Additionally, in Fig. \ref{figSeparation}, there is a region where $\hat \Phi^+ \cdot \hat \Phi_0 > 0.95$ (bounded by yellow dashed line). In this region,  $D<0$ and the $m=0$ projection is energetically favorable; the spin-1 system separates into a spinless non-magnetizable sector $\hat\Phi_0$, and a spin-$\frac{1}{2}$ sector exhibiting standard SC separation into the fields $\hat\Phi_s$ and $\hat\Phi_{\tilde c} = \frac{1}{\sqrt 2}(\hat \Phi_1 + \hat\Phi_{-1})$.  Figure \ref{figSeparation} shows that this region has a small overlap with the region of approximate QSC separation, so these two ways of decoupling the fields are sufficient for most of the phase space. In this work, we focus exclusively on the large QSC separated region.

Consequently, the bosonized BHH that shows the QSC separation is 
\begin{eqnarray}
\hat{H}&=& \sum_{\beta = q,s,c}\frac{u_\beta}{2}\int dx \left(\frac{1}{K_\beta}(\partial_x \hat \Phi_\beta)^2 +K_\beta\hat \Pi^2_\beta\right) \label{eqFullBosBHH}\\
&+& \mu_q\bar\rho^2 \int dx \ \cos(\sqrt{6\pi}\hat\theta_q ) \nonumber,
\end{eqnarray}
where $\beta$ indexes the separated sectors, $\mu_q = 2\Delta g/3$, the Luttinger parameters are $K_\beta = \pi \sqrt{t/U_\beta}$, and the self-interactions $U_\beta$ are
\begin{eqnarray}
U_s &=& \frac{\Delta g}{3} - D, \nonumber\\
U_c &=& \frac{3g}{2}-\frac{2D}{3} + \frac{11\Delta g}{9}, \label{eqInteractions} \\
U_q &=& -\frac{2\Delta g}{9} - \frac{D}{3}. \nonumber
\end{eqnarray}
While the functional form of $K_\beta$ is not quantitatively useful as previously discussed, the interaction strengths hold information about the behavior of the system; this will be explored in Sec. \ref{secSF}. Before that, let us focus on the symmetry point in which the spin and quadrupole sectors are free, and the interaction is due only to the charge. Here we change to the momentum basis, obtaining the linear holons $\hat a_i^{(c)}$ in the charge sector \cite{giamarchi}, and massive spinons $\hat a_i^{(s)}$ and quadruplons $\hat a_i^{(q)}$ in the magnetic sectors. The latter are free and therefore not LLs; however we have essentially rudimentarily applied a refermionization technique \cite{von_Delft_1998}, yielding
\begin{eqnarray}
\hat{H}&=& -2 t \sum_{k_{s}} \cos \left(k_{s}\right) \hat{a}_{k_{s}}^{(s) \dagger} \hat{a}_{k_{s}}^{(s)}\label{eqBHdiag}\\
&-&2 t \sum_{k_{q}} \cos \left(k_{q}\right) \hat{a}_{k_{q}}^{(q) \dagger} \hat{a}_{k_{q}}^{(q)} +u_c \sum_{p}|p| \hat{a}_{p}^{(c) \dagger} \hat{a}_{p}^{(c)}.  \nonumber
\end{eqnarray}
From this, we can determine the low-energy spectrum of the system, a cross-section $k_s=k_q$ of which is shown in Fig. \ref{figSpectrum}: free cosine bands in the magnetic sector separated by discrete holon excitations. Notice that the system becomes gapped at $3g = 32t$; this is a qualitative indication of the SF-MI transition \cite{fisher1989, Greiner2002}. Additionally, by comparing with Ref. \cite{manmana2011}, we find that our system becomes gapped when the physics is accurately captured by a Heisenberg-type Hamiltonian; we will explore this regime further in Sec. \ref{secHBB}.

\section{Magnetic phases of the superfluid}\label{secSF}

Generally, LLs are in the SF phase for $K_c>1$ \cite{giamarchi}; thus, let us study this regime with the Hamiltonian of Eq. \ref{eqFullBosBHH}. From the interactions of Eqs. \ref{eqInteractions}, we find three general behaviors, shown in Fig. \ref{figPhaseDiag}(a). First, when $\Delta g > 3D$ and $ 2 \Delta g < 3 D$, both magnetic sectors are repulsive and the whole system behaves like three decoupled LLs with $K_\beta \in \mathbb{R}$. It would seem like an opposite regime, in which $\Delta g < 3D$ and $ 2 \Delta g > 3 D$ and both magnetic sectors are attractive is possible; however, in this case vanishing compressibilities would require simultaneous saturation of the orthogonal spin and quadrupole densities. This being physically unreasonable leads us to believe the system must saturate only one of the densities at a time. Since for $\Delta g < 0$ ($\Theta < -3\pi/4$ or $g_2 < g_0$) the $S=2$ quintuplet interaction channel is more energetically favorable, and viceversa, we determine the transition between a spin-saturated magnetizable phase and a quadrupole-saturated non-magnetizable (Q) phase to take place at the $\Delta g = 0, \ D>0$ line.

In the magnetizable side, spin density saturates whereas the quadrupole sector is a LL. However, close to the symmetry point we have a small $\mu_q$, and expect $\hat \Phi_q$ to be weakly repulsive yielding a large $K_q$. Through standard renormalization group techniques \cite{giamarchi}, we find that, under these conditions, the phase field $\hat \theta_q$ is trapped in one of the minima of the cosine, which means the quadrupolar density is completely disordered. As such, we determine this magnetizable side to be in a ferromagnetic (FM) phase. 

On the other side, quadrupolar density saturates whereas the spin sector is a LL. Therefore, to characterize this side, we compute the following bosonized operators:
\begin{eqnarray}
\hat \rho_0 &=& \frac{1}{\sqrt{3\pi}}\hat\Phi_c + 2\bar\rho\cos\bigg(2\pi\bar\rho x + 2\sqrt{\frac{\pi}{3}}\hat\Phi_c\bigg)  \\
&-& 2\bar\rho\sin\bigg(2\pi\bar\rho x + 2\sqrt{\frac{\pi}{3}}\hat\Phi_c\bigg)  \nonumber\\
(\hat S^+)^2 &=&2\bar \rho \bigg[\cos\bigg(4\pi\bar\rho x + 4\sqrt{\frac{\pi}{3}} \hat\Phi_c\bigg) \\
&-&\sin\bigg(4\pi\bar\rho x + 4\sqrt{\frac{\pi}{3}} \hat\Phi_c\bigg)\nonumber\\
 &+& \cos\bigg(4\sqrt{\frac{\pi}{2}} \hat\Phi_s \bigg)\bigg] e^{i\sqrt{2\pi}\hat\theta_s} \nonumber
\end{eqnarray}
with which we calculate the scaling dimension of the correlations:
\begin{eqnarray}
\left\langle \hat \rho_0(x)\hat \rho_0(y)\right\rangle &\propto & \left( \frac{\delta}{r}\right)^{2K_c/3} \\
\left\langle (\hat S^+(x))^2 (\hat S^-(y))^2\right\rangle &\propto& \left( \frac{\delta}{r}\right)^{\frac{8}{3}K_c} + \left(\frac{\delta}{r}\right) ^{4K_s-1/K_s}, \nonumber
\end{eqnarray}
where $\delta$ is the cutoff length, $r=|x-y|$, and the decay exponents determine the behavior of the system. Notice the absence of quadrupole contributions to the operators and correlations; this happens because the saturated quadrupolar density implies an ordering of the sector that is not describable by a LL, and thus contributes a constant to the expected values. The operators are chosen by inspecting the possible quadrupolar states: $\hat \rho_0$ and $(\hat S^\pm)^2$ are order parameters to possible quadrupolar phases with nematic vectors along the axis or on the plane, respectively.

The susceptibilities can be computed from the corresponding correlations; we find that for $K_c>3$ both have divergent peaks meaning that an infinitesimal external coupling would order the fields. In other words, only the strong quantum fluctuations induced by the low dimensionality are stopping the system from ordering \cite{giamarchi}. In this case, the system adopts a singlet SF quasiorder. On the other hand, if $1<K_c<3$, only the planar susceptibility diverges and the system becomes a planar nematic (XYN) SF. Our findings for the SF phase diagram are compatible with previous studies of similar systems \cite{GarciaRipoll2004, bulhakov, toth2012}.

\section{Magnetic phases of the Mott insulator}\label{secHBB}

\begin{figure}
\centering
\includegraphics[width = \linewidth]{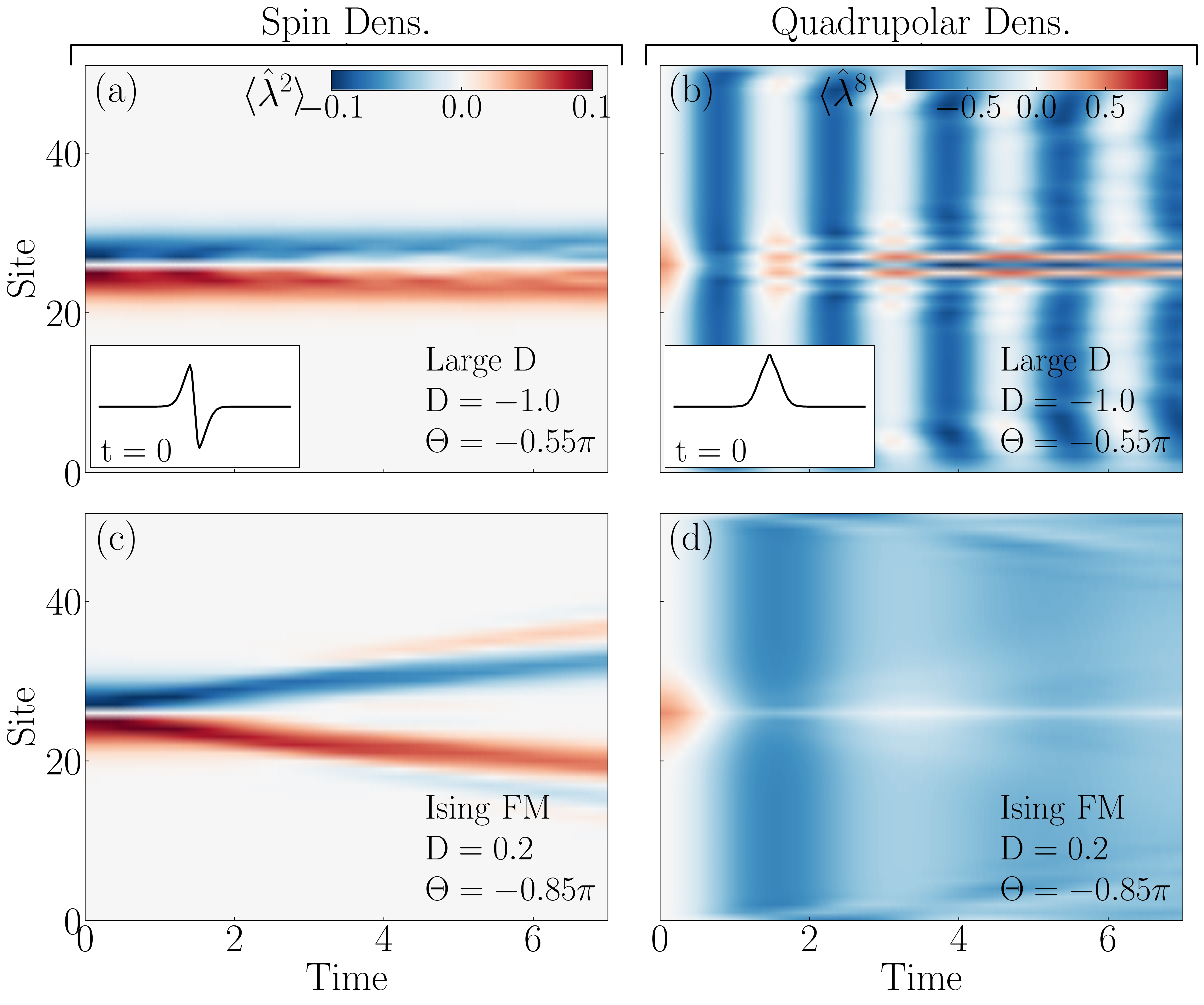}
\caption{\label{figNumerical} Time evolution of the MI using the time-MPS algorithm. (a-b) show the quadrupole density oscillating rapidly while spin fluctuations remain static in the large-$D$ phase. (c-d) show the opposite: propagating spin fluctuations with a stable quadrupole sector in the Ising FM phase. Since charge is frozen out of the MI, these results show QSC separation away from the symmetry point.}
\end{figure}

For $K_c=1$, the system becomes a single-filled MI \cite{Greiner2002}, the charge sector is frozen out, and the degeneracy of the magnetic sectors is partially lifted by the interaction and the quadratic Zeeman field, unfolding variety of magnetic phases. Although the Hamiltonian of Eq. \ref{eqFullBosBHH} does not fully capture the effect of a deep-well lattice, we can describe the MI with the Heisenberg bilinear-biquadratic (HBB) model \cite{imambekov2003}
\begin{equation}
\hat{H}=-\sum_{i}J_1\hat{\mathbf{S}}_{i} \cdot \hat{\mathbf{S}}_{i+1}+J_2\left(\hat{\mathbf{S}}_{i} \cdot \hat{\mathbf{S}}_{i+1}\right)^{2}+D\left(\hat{S}_{i}^{z}\right)^{2}, \label{eqHBB}
\end{equation}
obtained from the BHH of Eq. \ref{eqBHH} through second order quasi-degenerate pertubation theory on the hopping term \cite{Winkler2003}. In the standard parametrization, $J_1 = \cos\Theta$, and $J_2 =\sin\Theta$.

The HBB model can be rewritten \cite{bulhakov, peletminski, toth2012} in terms of the eight $\hat \lambda_i^\alpha$ generators at site $i$ as $\hat H = \sum_{i,\alpha} J_\alpha \hat \lambda_i^\alpha \hat \lambda_{i+1}^\alpha$, where $J_{\alpha} =\cos(\Theta) - \sin(\Theta)/2$ for $\alpha=2,5,7$, and $J_\alpha = \sin(\Theta)/2$ else. Notice that at $\Theta =-3\pi/4$ all generators contribute equally, manifesting the above-mentioned full $SU(3)$ symmetry; at this point the HBB model is equal to the Lai-Sutherland model \cite{piroli2019}. 

In order to show the QSC separation explicitly, we numerically simulated the time evolution of the HBB model using the time-matrix product states (time-MPS) algorithm~\cite{Vidal2003a,Verstraete2004}. Figure \ref{figNumerical} shows that for initial conditions with localized spin and quadrupole fluctuations (shown in insets), the sectors evolve independently: in the FM side (c-d), spin fluctuations propagate through the system while the quadrupole sector remains relatively stable. Contrariwise, in the Q-regime (a-b), the spin sector is immobile while the quadrupolar density oscillates rapidly. Additionally, by simulating the system in the MI regime, we automatically have charge separation as this degree of freedom remains static.

\begin{figure*}
\centering
\includegraphics[width=\linewidth]{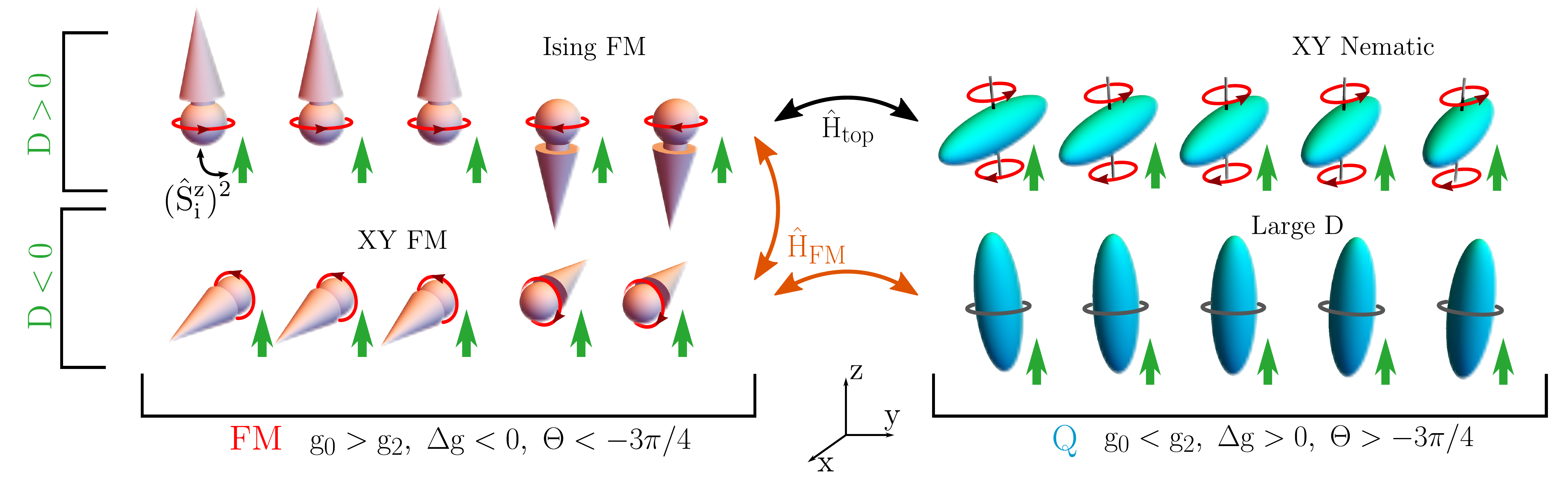}
\caption{\label{figPhases} Graphical depiction of the magnetic phases in the MI regime. The arrows denote the effective Hamiltonians (defined in text) that describe the particular transitions. In the FM side, the magnetizable spinor states can order along the axis or quasiorder on the plane. In the AF side, the nematic axes quasiorder on the plane or order along the axis. For simplicity, the quadrupolar states are shown as ellipses ellongated along the nematic axis.}
\end{figure*}

Although the HBB model is complicated to encompass, we use the general magnetic behavior extracted from Eq. \ref{eqFullBosBHH} to derive effective Hamiltonians made up of the most relevant terms for a certain region of phase space. Then, these effective Hamiltonians are bosonized and analyzed through the techniques mentioned in Sec.~\ref{secOverBos}, particularly the Luttinger parameter and the decay of the correlations. The results of the following discussion are summarized schematically by Fig. \ref{figPhases}, which shows the phases found and how they are connected by the effective Hamiltonians.

\textit{For $\Theta<-3\pi/4$}, we showed that the system is in a FM phase due to low energy cost of the quintuplet states. Since the spin sector dominates, the low-energy properties are captured by the effective Hamiltonian
\begin{equation}
\hat{H}_{F M}=-J_{t} \sum_{i}\hat{\mathbf{S}}_{i} \cdot \hat{\mathbf{S}}_{i+1}+D\left(\hat{S}_{i}^{z}\right)^{2}, \nonumber
\end{equation}
where $J_t = \cos(\Theta) - \sin(\Theta)/2$, and the quadrupolar terms are neglected. This model can be bosonized (as done by Schulz in Ref. \cite{schulz}) by projecting the spin-1 states into the triplet of two spin-$\frac{1}{2}$, $\hat{\boldsymbol{\sigma}}$ and $\hat{\boldsymbol{\tau}}$, which are subsequently mapped into fermions through the Jordan-Wigner (JW) transformation (see below) \cite{jordanWigner}. The fermions can be bosonized as discussed in Sec. \ref{secOverBos}, yielding a decoupled Hamiltonian $\hat H_{FM}=\hat H_f + \hat H_a$, with
\begin{eqnarray}
\hat{H}_{f} &=&\frac{J_{t}}{2} \int \mathrm{d} x\left(K_{f}^{2}\left(\partial_{x} \hat{\Phi}_{f}\right)^{2}+\left(\hat{\Pi}_{f}\right)^{2}\right)\\
&+& \bar\mu_{1} \int \mathrm{d} x \cos \left(\sqrt{8\pi} \hat{\Phi}_{f}\right)\nonumber \\
\hat{H}_{a} &=&\frac{J_{t}}{2} \int \mathrm{d} x\left(K_{a}^{2}\left(\partial_{x} \hat{\Phi}_{a}\right)^{2}+\left(\hat{\Pi}_{a}\right)^{2}\right)\\
&+& \bar\mu_{2} \int \mathrm{d} x \cos \left(\sqrt{8\pi} \hat{\Phi}_{a}\right) + \bar\mu_{3} \int \mathrm{d} x \cos \left(\sqrt{2\pi} \hat{\theta}_{a}\right) \nonumber
\end{eqnarray}
where $\hat \Phi_f = (\hat{\Phi}_\sigma + \hat{\Phi}_\tau)/\sqrt{2}$ and $\hat \Phi_a = (\hat{\Phi}_\sigma - \hat{\Phi}_\tau)/\sqrt{2}$ are the fields that diagonalize the system, $K_f^2 = 1-\frac{2}{\pi}(3+D)$, $K_a^2=1-\frac{2}{\pi}(1-D)$, $\bar\mu_i = J_t\mu_i/(\pi a)^2$, $\mu_1=\mu_2=1-D$, and $\mu_3=-\pi$. $\hat \Phi_f$ is a magnetizable field, due to being the sum of the spin-$\frac{1}{2}$, whereas $\hat \Phi_a$ is the opposite. This indicates a ferromagnetic (FM) region and a large-$D$ region, dominated by the $m=0$ projection, in this side of the phase diagram. The Luttinger parameters allow us to place them more precisely: $K_f$ becomes imaginary for large values of $D$, so that an Ising FM phase forms, characterized by axial domains of $|1\rangle$ or $|-1\rangle$. On the opposite end, large negative values of $D$ order $\hat \Phi_a$, indicating a large $D$ phase. In between, only the $\cos(\sqrt{2\pi}\hat\theta_a)$ is relevant, trapping the phase in a minimum of the cosine. As in Sec. \ref{secSF}, information is extracted by computing the bosonized operator
\begin{eqnarray}
\hat S^+_n &=& e^{i\sqrt{\pi/2}\hat\theta_f} \Big( \cos\big(\sqrt{\pi/2}\hat\theta_a\big) \\
&+& (-1)^n e^{i\sqrt{2\pi}\hat\phi_f}\cos\big(\sqrt{2\pi}\hat\phi_a + \sqrt{\pi/2}\theta_a\big) \Big) \nonumber
\end{eqnarray}
and the corresponding correlation
\begin{equation}
\left\langle \hat S^+_n \hat S^-_0 \right\rangle \propto \frac{1}{r^{1/4K_f}}.
\end{equation}
In this area, $K_f$ is small, meaning that there is a planar quasiorder and an XYFM phase due to the competition between the interaction-induced magnetization and the field that favors $|0\rangle$. The transition between Ising FM and XYFM phases is known to be at the isotropic line $D=0$ \cite{schulz, Franchini2017}.

\textit{For $\Theta>-3\pi/4$}, we showed that the system is in a non-magnetizable Q-regime. For large negative values of $D$, we expect a large-$D$ phase, since the quadratic Zeeman field makes $|\pm 1\rangle$ too costly. However, in the upper part of the phase diagram ($D>0$), these magnetizable states are favorable, and there is a competition between the field that prefers the axis, and the interaction which prefers a singlet. To capture this dynamic, we believe that the relevant generators are $\hat\lambda^2_i = \hat S^z_i$, and $\hat\lambda^1_i = -i ((\hat S^+_i)^2 - (\hat S^-_i)^2)/\sqrt{2}$ and $\hat\lambda^3_i = - ((\hat S^+_i)^2 + (\hat S^-_i)^2)/\sqrt{2}$ which are clearly related to the XYN order parameters from Sec. \ref{secSF}. Notice that these generators form an $SU(2)$ subalgebra; since the $m=0$ projection is not relevant in this region of the phase diagram, we can treat these three operators as Pauli matrices, and formulate a second effective Hamiltonian
\begin{equation}
\hat{H}_{\rm top}=\sum_i J_{p}\left(\sigma_{i}^{x} \sigma_{i+1}^{x}+\sigma_{i}^{y} \sigma_{i+1}^{y}\right)+J_{t} \sigma_{i}^{z} \sigma_{i+1}^{z},
\end{equation}
in terms of the spin-$\frac{1}{2}$ operators $\hat \sigma^z_i$, and $\hat \sigma^\pm_i$, which correspond to the spin-1 operators $\hat S^z_i$ and $(\hat S^\pm_i)^2$. In other words, the top of the phase diagram is described by an effective Heisenberg XXZ model \cite{Franchini2017, giamarchi}, where the planar coupling is $J_p = \sin(\Theta)/2$ and the axial anisotropic coupling is $J_t = \cos(\Theta)-\sin(\Theta)/2$. Through the JW transformation, the Hamiltonian is mapped onto interacting fermions, $\hat H_{\rm top} = -J_p \sum_i( \hat c^\dagger_i\hat c_{i+1} + {\rm hc}) + 2\Delta\sum_i \hat c^\dagger_i\hat c^\dagger_i\hat c_{i}\hat c_{i}$, where $\Delta= J_t/J_p$. Bosonizing as in Sec. \ref{secOverBos}, we retrieve the LL Hamiltonian of Eq. \ref{eqLuttFields}, with $K=\sqrt{1-4\Delta/\pi}$.

Again, we compute the bosonized operators
\begin{eqnarray}
\hat \sigma^-_n &=& e^{i \sqrt{\pi} \hat{\Phi}} ((-1)^n \hat{\psi}_{R}(x)+\hat{\psi}_{L}(x))\\
\hat \sigma^z_n &=& \frac{1}{\sqrt\pi}\partial_x\hat\Phi + \frac{(-1)^n}{\pi a}\cos(2\sqrt{\pi}\hat\Phi(x)),
\end{eqnarray}
and the corresponding axial and planar correlations \cite{affleckFields}
\begin{eqnarray}
\langle \hat \sigma^+_n \hat \sigma^-_0 \rangle &\propto& (-1)^{n} \frac{1}{r^{2 K+ 1/2K}} \\
\langle \hat \sigma^z_n \hat \sigma^z_0 \rangle &\propto& (-1)^{n} \frac{1}{2 \pi^{2}r^{2 K}}.
\end{eqnarray}
From the scaling dimensions, we find that for $\Delta$ close to zero, which happens inside the Q-regime, the system has a planar quasiorder; in terms of the original spin-1 operators this translates to an XYN phase. On the contrary, if $\Delta$ increases to a large value, entering the FM side, $K$ becomes imaginary and the Ising FM phase is retrieved \cite{schulz, sachdev_2011}. The spin chain is once again critical at the isotropic line $\Delta=1$, or $\Theta = -3\pi/4$, where the transition is found.

Putting everything together, we obtain the phase diagram shown in Fig. \ref{figPhaseDiag}(b), where the undetermined transition between the XYFM and large-$D$ phases is depicted as a blue shaded area. All the transition lines must converge at the critical $SU(3)$ point, where all the generators contribute equally, and our effective Hamiltonian approach breaks. Furthermore, there is an additional phase between the XYN and large-$D$: the dimer phase \cite{Rodriguez_2011, Xu2021}, characterized by entangled neighboring spins that take on the singlet state in order to minimize magnetization. The high entanglement makes it unsuitable for bosonization \cite{latorre2004, vidal2003, ding2012, santos2013}, so other approaches, such as Refs. \cite{Rodriguez_2011, kolezhuk2008}, are more appropriate to capture this phase.

\section{Conclusions}\label{secConclusion}

Using an effective bosonized field theory and numerical time-MPS simulations, we find a novel QSC separation in a gas of spin-1 bosons, showing its robustness against perturbations around the symmetry point, and characterizing it appropriately in terms of physical quantities. Not only does this allow us greater physical insight, but builds upon previous investigations into the role of the quadrupolar degrees of freedom in spin-1 systems \cite{bulhakov, peletminski, toth2012} and it reinterprets the scarce but valuable existing studies on similar separations. Additionally, Fig. \ref{figSeparation} can inform future numerical explorations of the problem, since the absence of an appropriate decoupled basis can hinder computations on the Q-regime.

Through the single-particle spectrum we qualitatively demonstrated the existence of the phase transition between the SF and the MI and determined when the system can be described by a Heisenberg-like Hamiltonian, validated by previous work \cite{manmana2011}. Bosonizing the BHH, we found the SF phase diagram, divided into FM, Q, and LL phases, which are compatible with similar previous studies \cite{GarciaRipoll2004, bulhakov}; our results differ by taking into consideration the quadratic Zeeman field and inspecting the Q phase more precisely, finding singlet and nematic behavior. For the MI, we decomposed the HBB into different effective Hamiltonians based on the general behavior of the BHH we found. Bosonizing the effective Hamiltonians, we found four out of five phases that exist, as shown by other studies like that of Rodr\'iguez \textit{et al} \cite{Rodriguez_2011}. 

Our methods provide an accessible way to analytically study spinor gases without resorting to common mean field methods that may become inadequate in 1D due to the strong collective quantum fluctuations. Through this approach, we were able to specify the emerging behaviors with considerable qualitative detail within the radically different SF and MI regimes. Furthermore, although advanced generalized forms of bosonization exist \cite{Witten1984, affleck1986, huang2021} and have been used to study the HBB model \cite{inami1992, allen2000}, these approaches do not focus on the phase diagram while posing a greater technical challenge; contrariwise, we keep our techniques flexible and accessible to a wide range of backgrounds and obtain a good characterization of the different phases. Thus, we expect our approach to be applicable to other interesting and important systems of condensed matter physics.

\section{Acknowledgements}

This work has been supported by Universidad del Valle under the internal project Cl. 71270.

\bibliography{biblio}

\end{document}